\documentclass[aip,jap,reprint]{revtex4-1}

\usepackage[version=3]{mhchem}
\usepackage{amsmath}
\usepackage{mathtools,amssymb}
\usepackage{graphicx}
\usepackage{natbib}
\usepackage{verbatim}
\usepackage[withbib, all]{authorindex}	
\usepackage[usenames]{color}
\usepackage{bm}
\usepackage[bookmarks=false]{hyperref}
\usepackage{siunitx}
\DeclareSIUnit\sq{\ensuremath{\Box}}
\DeclareSIUnit\mT{\milli\tesla}

\hypersetup{
   colorlinks, linkcolor={blue},
    citecolor={blue}, urlcolor={blue},
} 

\begin{document}

\title{Correlating spin transport and electrode magnetization in a graphene spin valve: simultaneous magnetic microscopy and non-local measurements}

\author{Andrew J. Berger}
\affiliation{Department of Physics, The Ohio State University, Columbus, OH 43210, USA}
\email{berger.156@osu.edu}
\author{Michael R. Page}
\affiliation{Department of Physics, The Ohio State University, Columbus, OH 43210, USA}
\author{Hua Wen}
\affiliation{Department of Physics, The Pennsylvania State University, University Park, PA, USA}
\author{Kathleen M. McCreary}
\affiliation{United States Naval Research Laboratory, Washington, D.C. 20375, USA}
\author{Vidya P. Bhallamudi}
\affiliation{Department of Physics, The Ohio State University, Columbus, OH 43210, USA}
\author{Roland K. Kawakami}
\affiliation{Department of Physics, The Ohio State University, Columbus, OH 43210, USA}
\affiliation{Department of Physics and Astronomy, University of California, Riverside, CA 92521, USA}
\author{P. Chris Hammel}
\affiliation{Department of Physics, The Ohio State University, Columbus, OH 43210, USA}
\email{hammel@physics.osu.edu}



\newpage
\begin{abstract}

Using simultaneous magnetic force microscopy (MFM) and transport measurements of a graphene spin valve, we correlate the non-local spin signal with the magnetization of the device electrodes. The imaged magnetization states corroborate the influence of each electrode within a one-dimensional spin transport model and provide evidence linking domain wall pinning to additional features in the transport signal.

\end{abstract}


\maketitle

Electrical injection and detection of spin have become common techniques to study spin transport and relaxation in metals and semiconductors \cite{johnson_interfacial_1985, hanbicki_efficient_2002, lou_electrical_2007, tombros_electronic_2007, appelbaum_electronic_2007, dash_electrical_2009} and are attractive for technological implementation of spin-based logic.  Graphene has demonstrated great promise as the spin transport channel in such devices, on account of its relatively long spin diffusion length and lifetime at room temperature \cite{tombros_electronic_2007, han_electrical_2009, han_tunneling_2010}.  Optimization of these devices necessitates an understanding of the fundamental interactions governing spin transport and will require a variety of measurement tools and techniques.  The successful combination of scanning probe microscopy with charge transport measurements \cite{topinka_imaging_2000, topinka_coherent_2001,berezovsky_imaging_2010,martin_observation_2008, jalilian_scanning_2011,panchal_visualisation_2014} suggests great opportunities for similar integration with spin transport devices. The imaging mode demonstrated in this letter---simultaneous MFM and spin transport---provides independent and complementary probes of magnetization, and facilitates a more complete understanding of spin and magnetization coupling in nanoscale devices. This approach provides a useful tool for studying complex magnetization configurations and their relationship to transport in other technologically relevant devices \cite{hayashi_current-controlled_2008}. 

Here we report direct imaging of electrode magnetization of an operating non-local graphene spin valve.  These images allow unambiguous determination of the overall magnetic state of the device and can be correlated with simultaneously acquired non-local magnetoresistance.  Using this correlation, we verify that a one-dimensional spin diffusion model \cite{costache_spin_2006}, combined with electrode magnetization switching, quantitatively captures the primary behavior of the device---high, low, and intermediate resistance states.  

Direct correlation of imaging and transport also provides insight into the origins of various observed features in the transport signal beyond simple binary switching of electrode magnetization.  We observe domain wall pinning and de-pinning in the ferromagnetic electrodes concurrent with otherwise obscure transitions in the non-local signal.  Without imaging, the origins of these occasionally-detected transitions \cite{tombros_electronic_2007} would remain unclear.  Our observation adds to a physical understanding of these sources of noise-like variations and aligns with the previous observation that ferromagnetic contacts can have a strong influence on spins in the transport channel via stray field effects \cite{dash_spin_2011-1}.    

The graphene spin valves used in this study consist of exfoliated single layer graphene with Co/MgO ferromagnetic contacts (see Ref. \citenum{han_spin_2012} for details). Using a combination of two-point and four-point measurements, we find the particular device described in this manuscript exhibits a graphene sheet resistance of \SI{1.38}{\kilo\ohm/\sq} and an average contact resistance of \SI{6.13}{\kilo\ohm}.

\begin{figure}[htp]
	\centering
		\includegraphics[width=\linewidth]{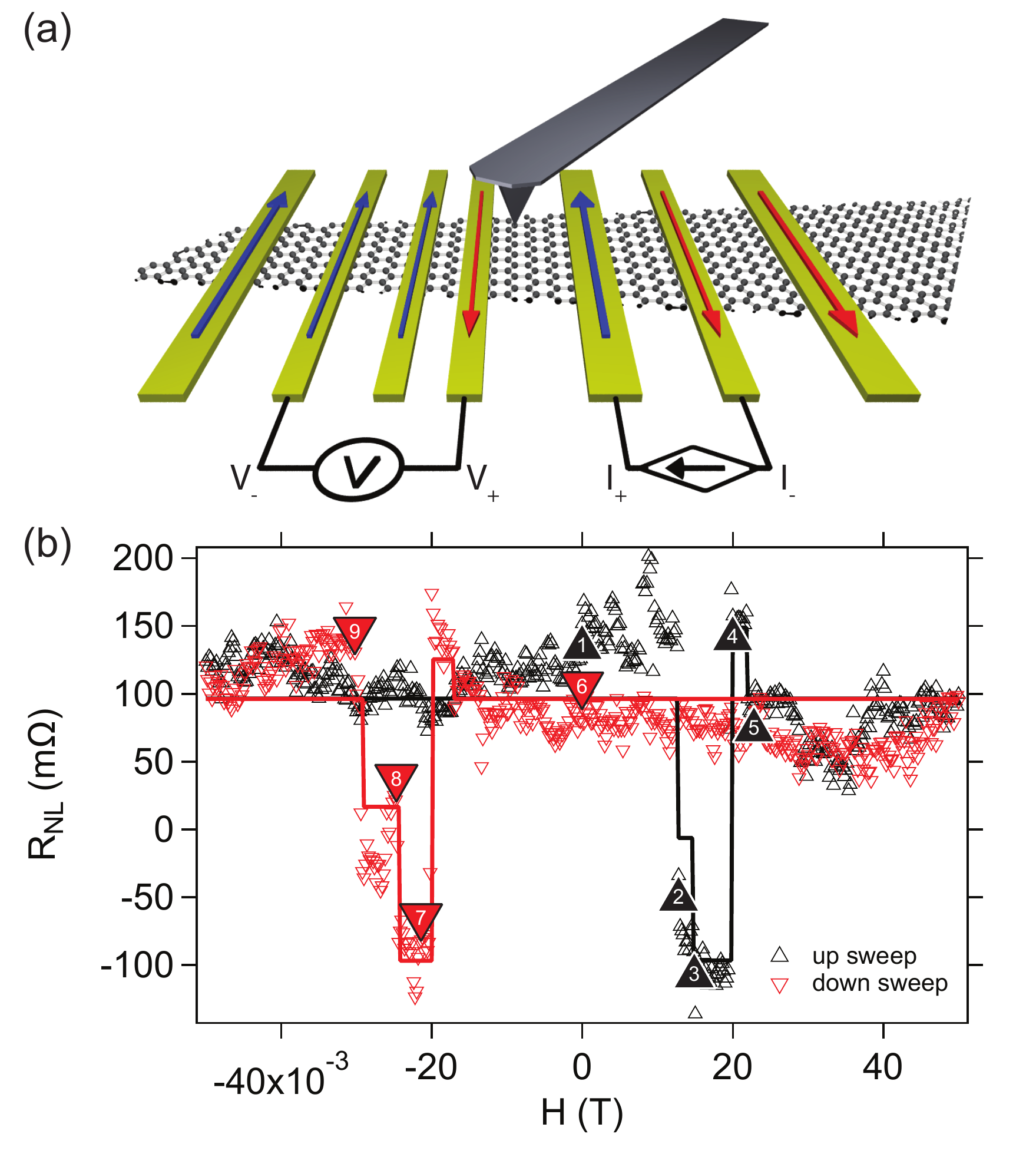}
	\caption{(a) Experimental schematic showing non-local circuit configuration of a graphene spin valve and the scanned magnetic probe positioned above the device.  The device contains 7 magnetic Co electrodes, 4 of which are used for transport measurements.  One possible magnetic configuration is shown with colored arrows representing magnetization direction.  (b) Electrically-detected non-local spin valve signal.  At magnetic field values indicated by numbered triangles, we acquired MFM images. A fit to the data is performed using the MFM-determined magnetic configuration of each state (solid lines).} 
	\label{fig:NLMR}
\end{figure}

\begin{figure}[htp]
	\centering
		\includegraphics[width=\linewidth]{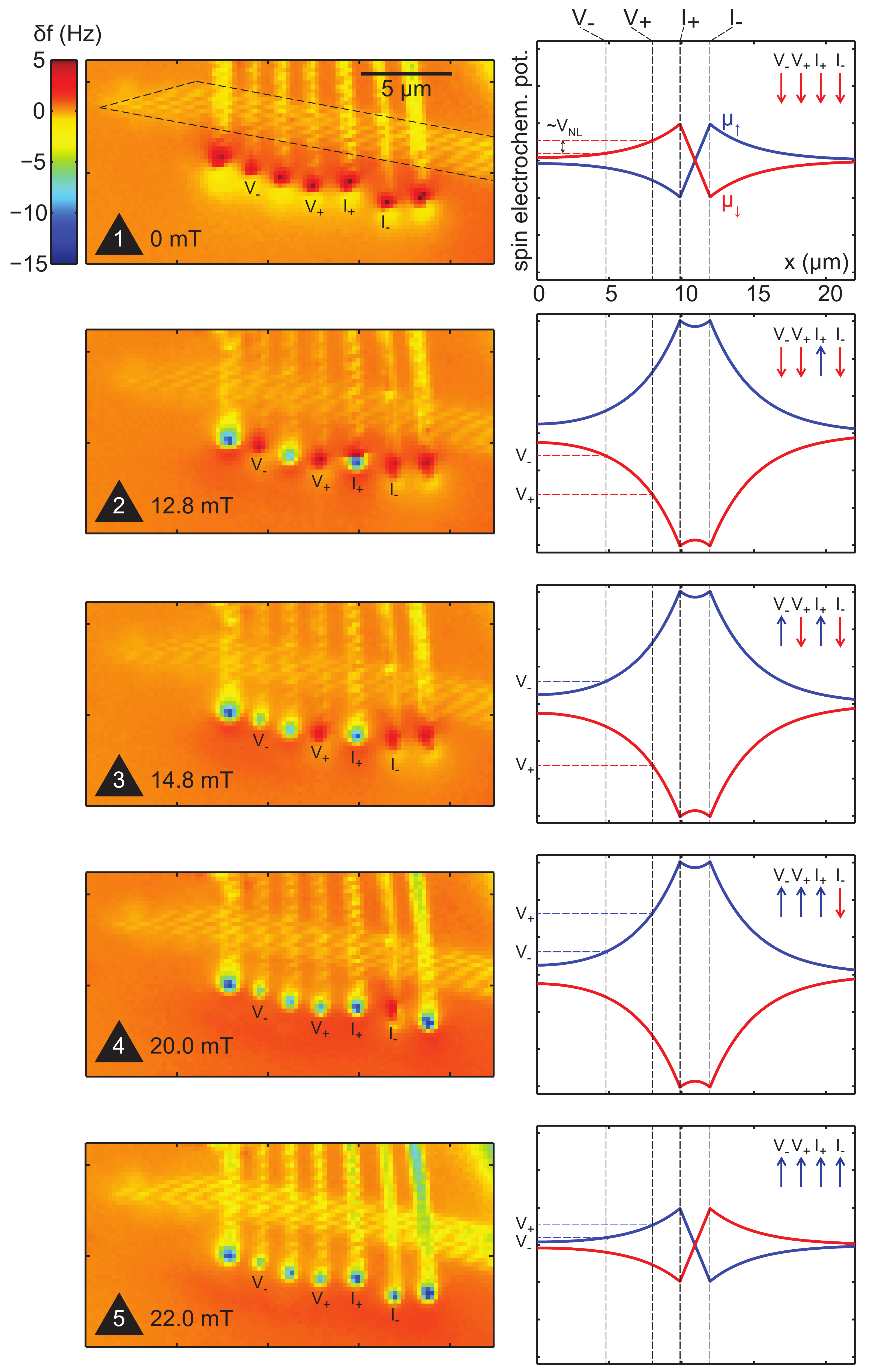}
	\caption{LEFT: MFM images acquired during magnetic field up sweep.  Color scale represents measured cantilever frequency shift, caused by tip-sample forces.  Images correlate with the $R_{\rm{NL}}$ measurements in Fig. \ref{fig:NLMR}(b) indicated by black numbered triangles.  RIGHT: 1-D spin diffusion calculation, showing spin-resolved electrochemical potential as a function of position in the graphene channel for up (blue) and down (red) spins.  Positions of the four circuit electrodes ($V_-, V_+, I_+$, and $V_-$) are indicated with dashed lines, and the overall device state (magnetization direction of electrodes) is indicated with up (blue) and down (red) arrows in the inset.  The measured voltage difference $V_{\rm{NL}}$ is proportional to the spin density difference $V_+ - V_-$ highlighted in these 1-D calculations.} 
	\label{fig:MFMUpSweep}
\end{figure}

\begin{figure}[htp]
	\centering
		\includegraphics[width = \linewidth]{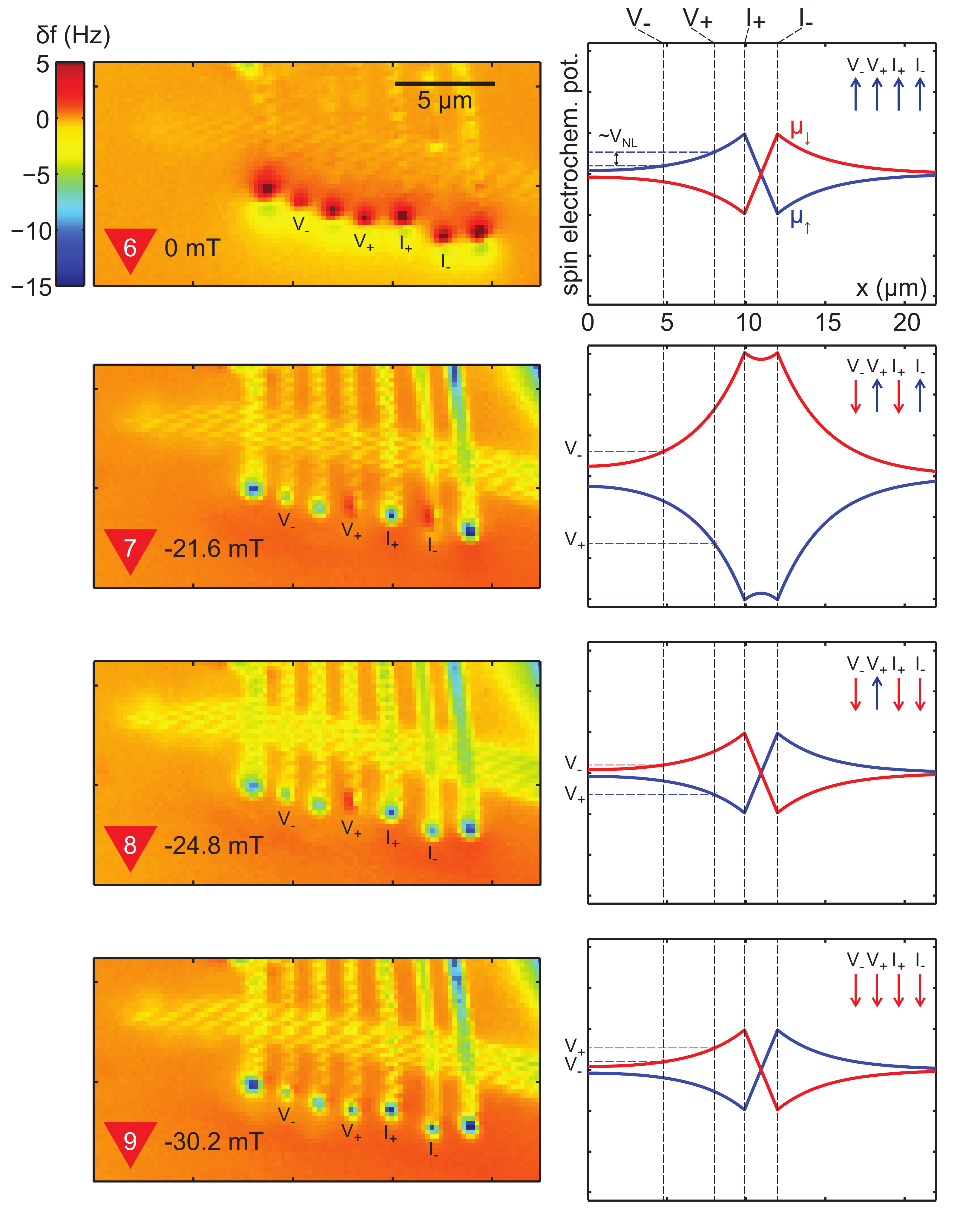}
	\caption{Sequence of MFM images and spin ECP profiles correlated with $R_{\rm{NL}}$ measurements at specific field values (red numbered triangles in Fig. \ref{fig:NLMR}(b)) acquired on magnetic field down sweep.  The MFM tip magnetization has reversed prior to imaging state 6, inverting the frequency shift color contrast relative to Fig. \ref{fig:MFMUpSweep}.  The electrode magnetizations are indicated by colored arrows in the spin density profile insets using the convention established in Fig. \ref{fig:MFMUpSweep}.} 
	\label{fig:MFMDnSweep}
\end{figure}

\begin{figure}[htp]
	\centering
		\includegraphics[width=\linewidth]{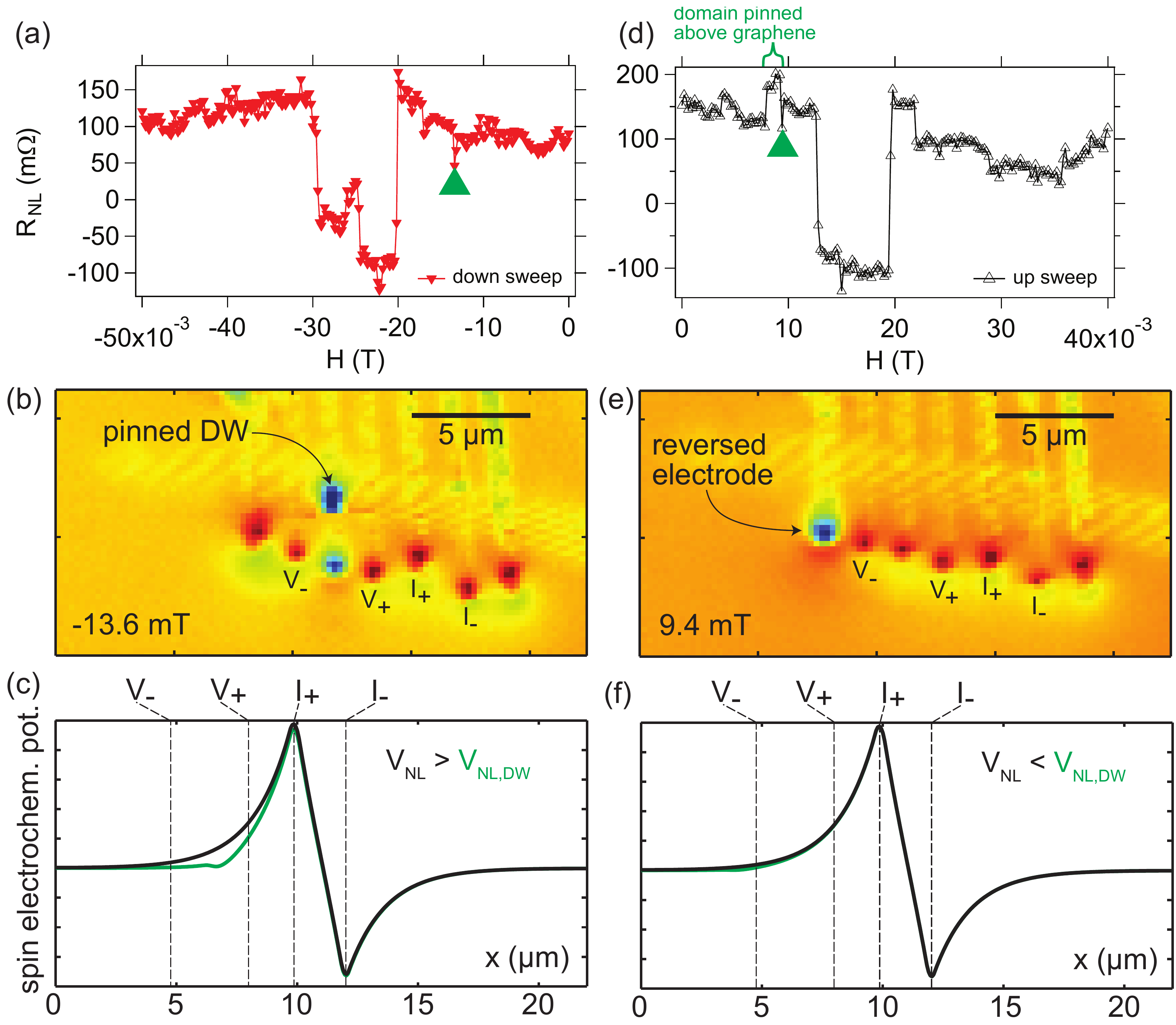}
	\caption{(a) Down-sweep NLMR signal.  Green triangle indicates field value at which the MFM image shown in (b) was acquired.  In this MFM image, a pinned domain wall can be clearly seen on the electrode between $V_-$ and $V_+$.  (c) Modeled spin ECP with ($V_{\rm{NL,DW}}$, green) and without ($V_{\rm{NL}}$, black ) the pinned domain wall.  The stray field from the domain wall causes local dephasing and a reduction in $V_{\rm{NL}}$ (green curve). (d) Up-sweep NLMR signal.  Green triangle indicates field value at which the MFM image shown in (e) was acquired.  In this MFM image, the outermost electrode has reversed (relative to the state 1).  Immediately before this electrode's magnetization reversed, the NLMR signal was seen to increase, and then return to baseline upon reversal.} 
	\label{fig:TDW}
\end{figure}

Spin transport measurements are performed using the traditional four-terminal non-local scheme \cite{tombros_electronic_2007, lou_electrical_2007} (see device schematic in Fig. \ref{fig:NLMR}(a)).  All measurements are performed at room temperature and in vacuum.  We source \SI{10}{\micro\ampere} of current at $I_+$ (drain at $I_-$) at \SI{11}{\hertz}, and measure the non-local voltage $V_{\rm{NL}} = V_+ - V_-$ with a lock-in amplifier.  Results are plotted as $R_{\rm{NL}} = V_{\rm{NL}}/I$.  An external magnetic field is swept in-plane, parallel to the long axis of the Co electrodes, in order to obtain the non-local magnetoresistance (NLMR) signal (Fig. \ref{fig:NLMR}(b)).  From a Hanle measurement of this device (with magnetic field oriented perpendicular to both the substrate plane and the injected spin orientation), we obtained a spin lifetime $\tau_s = $ \SI{476}{\pico\second} and spin diffusion length $\lambda_s = $ \SI{3.10}{\micro\meter} by using the measured device resistances and the fit algorithm of Ref. \citenum{sosenko_effect_2014}.  These parameters are used as inputs to the spin diffusion model to produce the spin electrochemical potential (ECP) profiles shown in Figs. \ref{fig:MFMUpSweep}, \ref{fig:MFMDnSweep}, and \ref{fig:TDW} and to fit our NLMR data (Fig. \ref{fig:NLMR}, solid lines).

We have previously described the custom-built scanning probe microscope which enables simultaneous acquisition of force microscopy and transport measurements \cite{berger_versatile_2014}.  This system utilizes frequency-shift detection of an oscillating cantilever to image atomic, electrostatic, and magnetic forces.  The magnetic force microscopy (MFM) signature of a bar magnet (which adequately approximates the Co electrodes) creates opposite frequency shifts at opposing ends of the bar, where the field gradient is strongest.  This MFM fingerprint enables clear determination of the magnetization orientation of each electrode.

At specific fields during the NLMR measurement field sweep (indicated by numbered triangles in Fig. \ref{fig:NLMR}(b)), we acquired MFM images of the graphene flake and the nearby ends of all seven Co electrodes.  Figs. \ref{fig:MFMUpSweep} and \ref{fig:MFMDnSweep} show the sequence of MFM images, constituting direct observation of magnetization orientation and switching sequence in an operational non-local spin valve.  These magnetization states can be explicitly correlated with the measured NLMR.  

To begin the measurement sequence, the field is first ramped to \SI{-250}{\mT} to initialize all electrodes to the parallel state.  The up sweep non-local voltage data is then recorded starting at \SI{-50}{\mT}.  The first MFM image is acquired on the up sweep at \SI{0}{\mT} applied field, verifying the all-parallel configuration (Fig. \ref{fig:MFMUpSweep}, subpanel 1).  The graphene flake is outlined in dashed lines for reference.  It is faintly observable in the force detection images owing to the periodic 11 Hz voltage applied to drive the injection current.  This sinusoidal voltage is experienced by the scanned cantilever as a periodic frequency shift due to the capacitive interaction between the grounded, conducting cantilever and the biased sample \cite{girard_electrostatic_2001}.  

On the up sweep, 5 distinct magnetic configurations are observed via MFM, corresponding to four unique $R_{\rm{NL}}$ levels (the all-parallel configurations in subpanels 1 and 5 exhibit the same resistance).  From this sequence of images, and correlation with the measured $R_{\rm{NL}}$ levels at the corresponding field, it is clear that all four circuit electrodes play critical roles in determining the measured non-local voltage.  The injector ($I_+$) and extractor ($I_-$) contacts determine the steady-state spin polarization in the channel, while the detector ($V_+$) and reference ($V_-$) contacts probe the spin chemical potential set by their respective magnetic orientations.  

A similar set of states are observed on the down sweep, as shown in Fig. \ref{fig:MFMDnSweep}.  The electrode magnetizations are again initialized to an all-parallel state with a large +\SI{250}{\mT} field.  This field reverses the MFM tip magnetization, which causes an inversion of the frequency shift contrast in Fig. \ref{fig:MFMDnSweep} relative to Fig. \ref{fig:MFMUpSweep}.  Therefore, the electrode magnetizations in states 5 and 6 are identical.  

\label{sec:1DModel}
With knowledge of the total magnetization state determined by the MFM images, we can model the spin ECP in the channel to understand the magnitude and sign of the measured non-local voltage.  The spin ECP spatial profile for each of the 9 observed electrode configurations is calculated using a 1-D spin diffusion model \cite{costache_spin_2006}.  These are shown immediately adjacent to the MFM images in Figs. \ref{fig:MFMUpSweep} and \ref{fig:MFMDnSweep} in order to highlight the correlation between the imaged magnetization orientation and the corresponding spin profile in the graphene.  Inputs to the model include the geometry of our device, the switching fields of each electrode as determined by MFM and transport, and the spin diffusion length (determined from a Hanle measurement).  The 1-D model does not include contact-induced spin relaxation \cite{maassen_contact-induced_2012} (i.e. once injected, graphene spins cannot escape through the metallic electrodes).  We find this to be justifiable given measured device resistances ($R_{\rm{contact}} > R_{\rm{graphene}}$) and the good agreement between experiment and fit.  Using this model, we can replicate the 6 unique voltage levels obtained in the electrically-detected non-local spin signal as shown from the fit in Fig. \ref{fig:NLMR} (solid lines).  There are no free parameters in the fit aside from a scaling of the overall magnitude, and a DC offset to match the measured data.  The relative voltage levels of the various resistance states are fixed by the experimentally-determined model inputs discussed above (i.e. the voltage level of each state cannot float freely with respect to the others).

The MFM images also uncover device behavior beyond the model of single domain switching.  We can correlate several transitions in the NLMR signal (Fig. \ref{fig:TDW}(a) and (d), green triangles) with interesting features in MFM images obtained at the same field.  This direct imaging provides evidence linking magnetization domain wall pinning and de-pinning to the observed changes in the NLMR signal.  Uncovering this connection between domain wall motion and spin transport is a key new capability made possible by simultaneous imaging and correlated transport.

Fig. \ref{fig:TDW}(b) shows a domain wall pinned directly above the graphene channel on the Co electrode between $V_-$ and $V_+$ at a field of \SI{-13.6}{\mT}.   This domain wall produces a large, but localized, stray field experienced by spins in the underlying graphene.  These spins are dephased by the field, causing a change in $R_{\rm{NL}}$ indicated by the green triangle in Fig. \ref{fig:TDW}(a).  Following the methods of Ref. \citenum{bhallamudi_imaging_2012}, we can model spin transport in the presence of such a localized, inhomogeneous field to obtain the new steady-state spin density.  Qualitatively, we  reproduce the observed reduction in spin signal.  We find that the dephasing has a greater impact on spins below $V_+$ than those under $V_-$, thereby reducing the voltage difference $V_{\rm{NL}}$ (see spin density profiles, Fig. \ref{fig:TDW}(c)). 
Growth of Co on graphene can result in enhanced grain and domain formation of Co due to graphene's high surface diffusion \cite{berger_magnetization_2014, han_spin_2012, zhou_deposition_2010}, and domains could be pinned above the graphene by these grain boundaries.

A similar feature is observed during the up sweep in Fig. \ref{fig:TDW}(d), between 8 and \SI{9.4}{\mT}.  At \SI{9.4}{\mT}, we acquired the MFM image in (e) showing that the leftmost electrode (external to the non-local circuit) had fully reversed relative to the \SI{0}{\mT} image of Fig. \ref{fig:MFMUpSweep}, subpanel 1.  Prior to this reversal, a domain wall necessarily propagated across the graphene channel which would have produced a stray field experienced by the graphene spins.  Modeling spin diffusion in the presence of this localized field reproduces the qualitative behavior of the electrically-detected NLMR: the non-local voltage is enhanced by the dephasing effect of the stray field (since spin accumulation near $V_-$ is primarily affected).  This enhancement persists until the electrode has completely reversed at \SI{9.4}{\mT} and the stray field has been eliminated.

In conclusion, we have utilized simultaneous transport and magnetic force microscopy to correlate the complete magnetization state of a non-local graphene spin valve with its spin signal.  We find good agreement between measured voltages and a 1-D spin diffusion model informed by the MFM images.  Such direct correlation studies also uncover device behavior that would be impossible to understand with transport measurements alone---namely, spin transport sensitivity to pinning and de-pinning of ferromagnetic domains in the contact electrodes.  

These results point to possible studies and applications in which spins interact with a much richer set of states than the two allowed orientations of a monolithically magnetized ferromagnet.  Understanding the interactions between spin transport and mobile domain walls or other magnetic textures in nanoscale geometries is an area of intense current research \cite{hayashi_current-controlled_2008}.  In-operando MFM of functioning devices is well suited for such studies.  For example, the fast and sensitive force detection provided by our microscope \cite{berger_versatile_2014} is able to detect domain wall transit below the scanned tip, which could be correlated with high-bandwidth features in transport data.  Furthermore, the local spin dephasing due to stray fields from the contact electrodes is identical to the spin imaging mechanism used in scanning spin precession microscopy \cite{bhallamudi_experimental_2013,bhallamudi_imaging_2012}, and suggests that spins in graphene should be amenable to such an imaging technique.  Overall, this approach shows clear promise for the precise characterization of spintronic device performance, and the development of reliable, application-ready devices.


Funding for this research was primarily provided by the Center for Emergent Materials, an NSF MRSEC at The Ohio State University (Award Number DMR-1420451).  V. B. acknowledges support from the Army Research Office under grant number W911NF-12-1-0587.

%

\end{document}